\documentstyle[12pt]{article}
\newcommand{\be}{\begin{eqnarray}}
\newcommand{\ee}{\end{eqnarray}}
\newcommand{\no}{\nonumber}
\newcommand{\add}{\addtocounter{equation}{+1}}
\newcommand{\addd}{\addtocounter{equation}{+2}}
\begin{document}
\begin{titlepage}
\vspace*{1.cm}
\begin{center}
{\Large\bf On the class of possible nonlocal anyon-like\\
operators and quantum groups}
\vskip 1.5 cm
by
\vskip 0.5 cm
{\bf M. Chaichian}$^{1,2}$, {\bf R. Gonzalez
Felipe}$^{1,}$\renewcommand{\thefootnote}{*}\footnote{ICSC-World Laboratory; On
leave of absence from
Grupo  de Fisica Te\'{o}rica, Instituto de Cibern\'{e}tica, Matem\'{a}tica
y Fisica, Academia de Ciencias de Cuba, Calle  E No. 309, Vedado,
La  Habana 4, Cuba.} {\bf and C. Montonen}$^3$
\end{center}
\vskip 2.0 cm

$^1$High Energy Physics Laboratory, Department of Physics,
P.O.Box 9 (Siltavuorenpenger 20 C), SF-00014 University of Helsinki,
Finland

$^2$Research Institute for High Energy Physics (SEFT),
P.O.Box 9 (Siltavuorenpenger 20 C), SF-00014, University of Helsinki,
Finland

$^3$Department of Theoretical Physics, P.O.Box 9 (Siltavuorenpenger 20 C),
SF-00014 University of Helsinki, Finland
\vskip 2.5 cm
{\bf Abstract:}
\vskip 0.5 cm
We find a class of nonlocal operators constructed by attaching a disorder
operator to fermionic degrees of freedom, which can be used to generate
$q$-deformed algebras following the Schwinger approach. This class includes
the recently proposed anyonic operators defined on a lattice.
\end{titlepage}
\noindent{\large\bf 1. Introduction}
\vskip 0.5 cm
Quantum groups and quantum algebras [1-3]
have attracted a great deal of
attention in recent years [4,5]. In particular, $q$-oscillators have been
formulated [6,7] and the Jordan-Schwinger approach [8] has been extensively
used for the constructions of quantum algebras [6,7,9] and quantum
superalgebras [10]. On the other hand, anyons [11] (particles with fractional
statistics) are of considerable interest since they appear in physical
systems. In recent papers [12-15], the construction of anyonic operators
on a two dimensional lattice has been studied. These anyonic oscillators
are substantially different from the $q$-oscillators, since the latters are
local operators and can be defined in any space dimension, while anyons are
intrisincally two dimensional nonlocal objects. (For a survey and discussion,
see e.g. [16]).

The anyonic operators on a lattice have been used to realize explicitly the
quantum algebra $SU_q(2)$ [15], as well as other $q$-deformed Lie algebras
[17], by means of a generalized Schwinger construction. The latter construction
differs from the one that involves $q$-oscillators in the sense that nonlocal
operators are used instead of local ones. Therefore, it is of interest to
study whether the anyons are the only nonlocal operators from which
$q$-deformed
algebras can be realized and whether there exists a more general class of such
operators.

In this letter we find a class of nonlocal operators, constructed by attaching
a disorder operator [12] to each fermionic degree of freedom, from which
the $q$-deformed algebras can be realized following the Schwinger approach.
Since this general class includes, as a particular case, the anyonic operators
defined on a lattice in [15], we shall call the operators in this class
"anyon-like".
\vskip 1.0 cm
\noindent{\large\bf 2. Bosonic and fermionic realization of quantum algebras}
\vskip 0.5 cm
It is well known that one of the simplest forms of the explicit realization
of $SU(2)$ algebra is the Schwinger construction [8]. This can be done by
introducing a pair of bosonic or fermionic oscillators. However, in the latter
case, unlike the bosonic one, only the 0 and 1/2 representations of $SU(2)$ are
recovered. To obtain the full set of representations, we must introduce several
copies of fermionic oscillators pairs.

Recently, the Schwinger construction has been also used to realize the
so-called $q$-deformed (quantum) algebras [6,7,9] and quantum superalgebras
[10]. If we introduce a pair of bosonic $q$-oscillators $a_i ,\ i=1,2$,
which satisfy the relations,
\be
& & a_ia^+_i-qa^+_ia_i=q^{-N_i}\ \ ,\ \ \ i=1,2\ ,\no\\
& & [a_i,a^+_j]=0\ \ ,\ \ \ i\neq j\ ,
\ee
then one can show that the generators defined as
\be
J^+=a^+_1a_2\ \ ,\ \ \ J^-=a^+_2a_1\ \ ,\ \ \ J^0=\frac{1}{2}(N_1-N_2)\ ,
\ee
satisfy the so-called $SU_q(2)$ algebra
\be
& & [J^0,J^\pm]=\pm J^\pm\ ,\no\\
& & [J^+,J^-]=[2J^0]\ ,
\ee
where
\be
& & [x]\equiv \frac{q^x-q^{-x}}{q-q^{-1}}\ .\no
\ee

We can use instead an ordered set of fermionic oscillators
$c_i(x)$, satisfying the anticommutation relations
\be
\biggl\{ c_i(x),c^+_j(y) \biggr\}=\delta_{ij}\delta(x,y)\ \ ,\ \ \
\biggr\{ c_i(x),c_j(y) \biggl\}=0\ ,
\ee
$i,j=1,2;$ where $x$ and $y$ belong to some countable , discrete set
$\Omega$, with an ordering relation defined on it, and $\delta(x,y)$ is a
delta function in $\Omega$\renewcommand{\thefootnote}{*}\footnote{This set
could be  a lattice when $x$ and $y$ are space coordinates.}. Then with
the use of the noncocommutative comultiplication [18], we define the local
(as functions of $x$) generators
\be
& &J^\pm(x)=\prod\limits_{y<x}q^{-j^0(y)}j^\pm(x)\prod\limits_{z>x}q^{j^0(z)}\
,
\no\\
& &J^0(x)=j^0(x)\ ,
\ee
where
\be
& &j^+(x)=c^+_1(x)c_2(x)\ \ ,\ \ \ j^-(x)=c^+_2(x)c_1(x)\ ,\no\\
& &j^0(x)=\frac{1}{2}\biggl\{ c^+_1(x)c_1(x)-c^+_2(x)c_2(x)\biggr\}\ ,
\ee
and
\be
& & [j^0(x),j^\pm(y)] = \pm j^\pm(x)\delta(x,y)\ ,\no\\
& & [j^+(x),j^-(y)] = 2j^0(x)\delta(x,y)\ .
\ee
{}From (5) and (7) it is straightforward to verify that
\be
& & [J^0(x),J^\pm(y)]=\pm J^\pm(x)\delta(x,y)\ ,\no\\
& & [J^+(x),J^-(y)]=0\ \ ,\ \ \ {\rm if}\ x\neq y\ ,\\
& & [J^+(x),J^-(x)]=\prod\limits_{y<x}q^{-2J^0(y)}\cdot
2J^0(x)\cdot\prod\limits
_{z>x}q^{2J^0(z)}\ ,\no
\ee
and that the global generators defined as
\be
J^\pm=\sum\limits_xJ^\pm(x)\ \ ,\ \ \ J^0=\sum\limits_xJ^0(x)\ ,
\ee
satisfy the $SU_q(2)$ algebra (3). Notice also that $(J^+(x))^+\neq J^-(x)$,
but we have instead the relation
\be
(J^+(x))^+=J^-(x)_{\left|\begin{array}{l}\ \\ q\rightarrow q^{-1}\end{array}
\right.}\ .
\ee

The above analysis, concerning the fermionic realization of $SU_q(2)$ algebra,
gives rise to the following question: do there exist operators, which can be
used to perform the Schwinger construction for the $SU_q(2)$ algebra (3)?
In ref. [15] it was shown that anyonic operators on a two-dimensional
lattice can be used to build this algebra explicitly. In the next section
we shall propose a more general class of nonlocal operators, which we shall
call "anyon-like" and which can be used in the Schwinger construction.
The use of the
terminology  "anyon-like" will be more clear when we define these
operators and obtain their commutation relations which are similar to those
of anyons.
\vskip 1.0 cm
\noindent{\large\bf 3. Schwinger construction from nonlocal anyon-like
operators}
\vskip 0.5 cm
Let us define the nonlocal operators $\alpha_i(x) ,\ \alpha^+_i(x)$ as
\be
& &\alpha_i(x)=q^{-\Delta_i(x)}c_i(x)\ ,\no\\
& &\alpha^+_i(x)=c^+_i(x)q^{\Delta_i(x)}\ ,
\ee
$(i=1,2)$, where $c_i(x),\ c^+_i(x)$ are the fermionic operators, which satisfy
(4) and
\be
& &\Delta_i(x)=\sum\limits_yf(x,y)N_i(y)\ ,\no\\
& &N_i(y)=c^+_i(y)c_i(y)\ ;
\ee
$x$ and $y$ belong to some discrete set $\Omega$ and $f(x,y)$ is an arbitrary
function to be specified later. The construction of the operators (11)
can be understood as a kind of Jordan-Wigner transformation [19] of the
fermionic operators, i.e., we attach the disorder operators [12]
$q^{-\Delta_i(x)}$ to the fermions $c_i(x)$.

{}From (12) and (4) it follows that
\be
& & [\Delta_i(x),c_j(y)]=-\delta_{ij}f(x,y)c_i(y)\ ,\no\\
& & [\Delta_i(x),c^+_j(y)]=\delta_{ij}f(x,y)c^+_i(y)\ \\
& & [\Delta_i(x),N_j(y)]=0\ .\no
\ee
Using (4) and (13) we can calculate the commutation relations of the operators
$\alpha_i(x),\ \alpha^+_i(x)$ defined in (11).
We have
$$\alpha_i(x)\alpha^+_i(y)+q^{f(y,x)-f(x,y)}\alpha^+_i(y)\alpha_i(x)=
\delta(x,y)\ ,\eqno{(14a)}$$
$$\alpha_i(x)\alpha_i(y)+q^{f(x,y)-f(y,x)}\alpha_i(y)\alpha_i(x)=0\ .
\eqno{(14b)}$$
Notice that when $x=y$, from (14) it follows that
\add
\be
\biggl\{\alpha_i(x),\alpha^+_i(x)\biggr\}=1\ ,\no\\
\alpha^2_i(x)=0\ ,\hspace{1.0cm}
\ee
i.e., the $\alpha_i,\ \alpha^+_i$ operators describe hard core objects which
obey standard (fermionic) anticommutation relations at the same point.
As a special case, the operators $\alpha_i,\ \alpha^+_i$ can be identified
with the anyonic operators on a lattice recently introduced in the
literature [12-15]. Indeed, the latter can be obtained from (11) by
taking $q=e^{i\nu\pi}$ and $f(x,y)=-\frac{1}{\pi}\theta(x,y)$, where $\nu$
is the statistics determining parameter and $\theta(x,y)$ is the lattice
angle function [13,20].

In what follows we shall refer to the operators $\alpha_i(x),\ \alpha^+_i(x)$
as
anyon-like operators, although it is clear that they are actually anyon
operators only for the particular choice of $q$ and $f(x,y)$ given above.

Our aim is to find the general class of functions $f(x,y)$ so that the
Schwinger construction can be performed using the operators (11). From the
analysis of the fermionic realization of the $SU_q(2)$ algebra given in the
previous section and, in particular, from relation (10) it follows
that the operators $\alpha_i(x),\ \alpha^+_i(x)$ are not sufficient
to construct the local generators $J^\pm(x)$. In order to construct them,
one can introduce an extra pair of operators $\tilde{\alpha}_i(x),\
\tilde{\alpha}^+_i(x),\ i=1,2$, defined as
\be
& &\tilde{\alpha}_i(x)=q^{\tilde{\Delta}_i(x)}c_i(x)\ ,\no\\
& &\tilde{\alpha}_i^+(x)=c^+_i(x)q^{-\tilde{\Delta}_i(x)}\ ,
\ee
with
\be
\tilde{\Delta}_i(x)=\sum\limits_yg(x,y)N_i(y)\  ,
\ee
where $g(x,y)$ is a function, whose properties will be determined later.
The commutation relations among the new operators (16) can be obtained
from eqs. (14),(15), by replacing $q$ by $q^{-1}$ and $f(x,y)$ by
$g(x,y)$.

It is worthwhile to give also the commutation relations between the
operators (11) and (16). We have
$$\alpha_i(x)\tilde{\alpha}^+_i(y)+q^{-(f(x,y)+g(y,x))}\tilde{\alpha}^+_i
(y)\alpha_i(x)=q^{-(\Delta_i(x)+\tilde{\Delta}_i(x))}\delta(x,y)\ ,
\eqno{(18a)}$$
$$\alpha_i(x)\tilde{\alpha}_i(y)+q^{f(x,y)+g(y,x)}\tilde{\alpha}_i(y)
\alpha_i(x)=0\ .\hspace{2.0cm}
\eqno{(18b)}$$
Let us define now the operators
\add
\be
& &J^+_\alpha(x)=\alpha^+_1(x)\alpha_2(x)\ ,\no\\
& &J_\alpha^-(x)=\tilde{\alpha}^+_2(x)\tilde{\alpha}_1(x)\ ,\no\\
& &J^0_\alpha(x)=\frac{1}{2}\biggl\{\alpha^+_1(x)\alpha_1(x)-\alpha^+_2
(x)\alpha_2(x)\biggr\}\\
&
&\hspace*{1.1cm}=\frac{1}{2}\biggl\{\tilde{\alpha}^+_1(x)\tilde{\alpha}_1(x)-\tilde{\alpha}
^+_2(x)\tilde{\alpha}_2(x)\biggr\}\ .\no
\ee
First notice that from the definitions (11) and (16) it follows that
\be
J^0_\alpha(x)=j^0(x)=\frac{1}{2}\biggl\{ N_1(x)-N_2(x)\biggr\}\ ,
\ee
where $j^0(x)$ is given in (6). Moreover, we can write
$$J^+_\alpha(x)=q^{-f(x,x)+2\sum\limits_yf(x,y)j^0(y)}j^+(x)\ ,
\eqno{(21a)}$$
$$J^-_\alpha(x)=q^{g(x,x)+2\sum\limits_yg(x,y)j^0(y)}j^-(x)\ ,
\eqno{(21b)}$$
with $j^\pm(x)$ also defined in (6).

Using relations (7), eqs.(21) can be rewritten in a more convenient form,
namely,
$$J^+_\alpha(x)=q^{(2j^0(x)+1)f(x,x)}\prod\limits_{y<x}q^{2f(x,y)j^0(y)}
j^+(x)\prod\limits_{z>x}q^{2f(x,z)j^0(z)}\ ,
\eqno{(22a)}$$
$$J^-_\alpha(x)=q^{(2j^0(x)+1)g(x,x)}\prod\limits_{y<x}q^{2g(x,y)j^0(y)}
j^-(x)\prod\limits_{z>x}q^{2g(x,z)j^0(z)}\ .
\eqno{(22b)}$$
Comparing eqs. (20), (22) with the local generators (5), we see that if we
define
\addd
\be
f(x,y)=g(x,y)=\left\{\begin{array}{ccc}-\frac{1}{2} & , & y<x\\ 0&,&y=x\\
\frac{1}{2}&,&y>x\end{array}\right.\ ,
\ee
then $J^\pm_\alpha(x)=J^\pm(x),\ J^0_\alpha(x)=J^0(x)$, and consequently,
the global generators defined as
\be
J^\pm_\alpha=\sum\limits_xJ^\pm_\alpha(x)\ \ ,\ \ \ J^0_\alpha=\sum\limits_x
J^0_\alpha(x)\ ,
\ee
will satisfy the $SU_q(2)$ algebra (3). It is clear that for any function
$f'(x,y)=af(x,y)=ag(x,y)$, where $a$ is a real constant, the operators (24)
will satisfy the algebra $SU_{q'}(2)$ with $q'=q^a$.

Let us remark that the choice of the functions $f(x,y)$ and $g(x,y)$ in
(23) in order to get the local generators (5) of the $SU_q(2)$ algebra,
is unique. In other words, there exists only one pair of operators
$\alpha_i(x),\ \alpha^+_i(x)$ defined by (11) and one pair of operators
$\tilde{\alpha}_i(x),\ \tilde{\alpha}_i^+(x)$ defined by (16), with $f(x,y)$
and $g(x,y)$ given in (23), so that the Schwinger construction can be performed
to build the local generators (5) of the $SU_q(2)$ algebra (3). On the other
hand, as we shall see below, there exists a more general class of functions
$f(x,y)$ and $g(x,y)$, such that the local operators (20) and (21) satisfy
the commutation relations (8) and in consequence, the global generators
(24) satisfy the $SU_q(2)$ algebra.

{}From eqs. (20), (21) and using (7) it is straightforward to verify that
\be
\ [J^0_\alpha(x),J^\pm_\alpha(y)]=\pm J^\pm_\alpha(x)\delta(x,y)\ ,
\ee
and
\be
J^+_\alpha(x)J^-_\alpha(y)-q^{-2(f(x,y)+g(y,x))}J^-_\alpha(y)J^+_\alpha(x)
\hspace{4.4cm}\no\\
\no\\
=q^{-(f(x,x)+g(x,x))+2\sum\limits_z(f(x,z)+g(x,z))J^0_\alpha(z)}
\cdot 2J^0_\alpha(x)\cdot\delta(x,y)\ .
\ee
Assume that for $\forall x,y\in\Omega$,
\be
f(x,y)=-g(y,x)\ .
\ee
Then from (26) we conclude that
\be
\ [J^+_\alpha(x),J^-_\alpha(y)]=0\ \ ,\ \ \ \forall x\neq y\ ,
\ee
and
\be
\ [J^+_\alpha(x),J^-_\alpha(x)]=q^{2\sum\limits_z(f(x,z)-f(z,x))J^0_\alpha(z)}
\cdot 2J^0_\alpha(x)\hspace{3.0cm}\no\\
\no\\
=\prod\limits_{y<x}q^{2(f(x,y)-f(y,x))J^0_\alpha(y)}\cdot 2J^0_\alpha(x)
\prod\limits_{z>x}q^{2(f(x,z)-f(z,x))J^0_\alpha(z)}\ .
\ee
{}From eq. (29) one sees that if the function $f(x,y)$ satisfies the relation
\be
f(x,y)-f(y,x)=\left\{\begin{array}{cccc}-\beta&,&{\rm for}&y<x\\
0&,&{\rm for}&y=x\\ \beta&,&{\rm for}& y>x\ ,\end{array}\right.
\ee
where $\beta$ is a real constant, and we define $q'=q^\beta$, then eq. (29) has
the same form as the one of the local generators of the $SU_{q'}(2)$ algebra
(compare with the last equation in (8)). Consequently, relations (25),
(28) and (29) will imply that the global generators (24) satisfy the
$SU_{q'}(2)$ algebra (3).

Thus we have found a wide class of nonlocal operators, namely, the
anyon-like operators (11) and (16), which generate the $SU_q(2)$ algebra
through the Schwinger construction. Any function $f(x,y)$ satisfying the
relation (30),which only restricts the antisymmetric part of $f$, yields
acceptable anyon-like operators. Our construction generalizes the anyon
operators of [15], and it would be interesting to consider generalizing
the construction even further.

\vskip 1.0 cm
\noindent {\large\bf 4. Example: Anyonic operators on a lattice}
\vskip 0.5 cm
As mentioned above (see our remark after eqs.(15)), a particular choice
of $q$ and $f(x,y)$ in eqs. (11) gives rise to the anyonic operators
defined on a two dimensional lattice [12-15]. For definiteness, we shall
assume that the lattice $\Omega$ has spacing one. In constructing the
anyonic operators, the basic element is the lattice angle function
$\theta(x,y)$ [13,19]. Here we shall use the specific description
of $\theta(x,y)$ given in [15]: The lattice $\Omega$ is embedded into a
lattice $\Lambda$ with spacing $\epsilon$. Then to each point $x\in\Omega$
one associates a cut $\gamma_x$, made with bonds of the dual lattice $\Lambda'$
from minus infinity to $x'=x+0$ along $x$-axis, with $0=(\frac{\epsilon}{2},
\frac{\epsilon}{2})$ the origin of $\Lambda'$. In the limit $\epsilon
\rightarrow 0$ we can endow the lattice with an ordering and define [15]
\be
\theta_{\gamma_x}(x,y)-\theta_{\gamma_y}(y,x)=\pi\ \  {\rm for}\ \ \ x>y\ ,
\ee
where $\theta_{\gamma_x}(x,y)$ is the angle of the point $x$ measured from
the point $y'\in\Lambda'$ with respect to a line parallel to the positive
$x$-axis. This definition of $\theta(x,y)$ is not unique since it depends
on the choice of the cuts. If we choose now for each point of the lattice
a cut $\delta_x$ made with bonds of the dual lattice $\Lambda'$ from plus
infinity to $x'=x-0$ along $x$-axis, we will have
\be
\tilde{\theta}_{\delta_x}(x,y)-\tilde{\theta}_{\delta_y}(y,x)=-\pi\ \ {\rm
for}\ \ \
x>y\ ,
\ee
where $\tilde{\theta}_{\delta_x}(x,y)$ is now the angle of $x$ seen from
$y'\in\Lambda'$ with respect to a line parallel to the negative $x$-axis.
Besides, we have also the relation
\be
\tilde{\theta}_{\delta_x}(x,y)-\theta_{\gamma_y}(y,x)=0\ \ ,
\ \ \ \forall x,y\in\Lambda\ .
\ee

The angle functions $\theta(x,y)$ and $\tilde{\theta}(x,y)$ allow to
distinguish between clockwise and counterclockwise braidings and, in
consequence, to introduce two types of anyonic operators $a(x_\gamma)$ and
$a(x_\delta)$, where $x_\gamma$  denotes the point $x\in\Omega$ with its
associated cut $\gamma_x$. The latter operators are defined as [15]
\be
& &a_i(x_\rho)=K_i(x_\rho)c_i(x)\ ,\no\\
& &K_i(x_\rho)=e^{i\nu\sum\limits_y\theta_{\rho_x}(x,y)N_i(y)}\ ,\\
& &N_i(y)=c^+_i(y)c_i(y)\ ,\no
\ee
where $\rho=\gamma_x$ or $\delta_x ,\ i=1,2$, $\nu$ is the statistical
parameter and $c_i(x)$ are the fermionic operators defined on $\Omega$,
which obey the anticommutation relations (4) with
$\delta(x,y)=1\  {\rm if}\  x=y ,\ \delta(x,y)=0\ {\rm if}\ x\neq y$.

{}From the definitions (11) and (16) for the $\alpha_i(x)$ and
$\tilde{\alpha}_i
(x)$ operators, and from the definitions (34) for the anyonic operators
$a_i(x_\gamma)$ and $a_i(x_\delta)$, we see that the latters can be obtained
from (11) and (16) by assuming
\be
& & q=e^{i\nu\pi}\  ,\no\\
& & f(x,y)=-\frac{1}{\pi}\theta_{\gamma_x}(x,y)\ ,\\
& & g(x,y)=\frac{1}{\pi}\tilde{\theta}_{\delta_x}(x,y)\ .\no
\ee

Now, relations $(31)-(33)$ will imply that eqs. (27) and (30) also hold, with
$\beta=1$ in (30). Thus, the global generators (24) defined through the local
generators (19), where now instead of $\alpha_i(x),\ \tilde{\alpha}_i(x)$
we have $a_i(x_\gamma),\ a_i(x_\delta)$, respectively, will satisfy the
$SU_q(2)$ algebra (3) with $q=e^{i\nu\pi}$.

It is interesting to remark that anyons can consistently be defined also on
a line (one dimensional chain). It that case one can define
\be
\theta_{\gamma_x}(x,y)=-\tilde{\theta}_{\delta_x}(x,y)=\left\{\begin{array}{ccc}
\frac{\pi}{2}&{\rm for}&x>y\\ \frac{-\pi}{2}&{\rm for}&x<y\ ,\end{array}
\right.\no
\ee
so that eqs. $(31)-(33)$ are fulfilled, Then, from (35) we obtain that
$f(x,y)=g(x,y)=-\frac{1}{2}$ for
$x>y$ and $\frac{1}{2}$ for $x<y$. If we assume now that $f(x,x)=g(x,x)=0$
(as in (23)) or, exclude the point $y=x$ from the definition of the disorder
operators (34), then we will have that the local generators (22) defined
through
the anyonic operators will coincide with the iterated coproduct (5) and in
consequence, the global generators (24) give the $SU_q(2)$ algebra with
$q=e^{i\nu\pi}$.
\vskip 2.0 cm
{\bf Acknowledgement}

One of us (R.G.F.) would like to thank the World Laboratory for financial
support.

\pagebreak

\end{document}